\documentclass[pra,twocolumn,superscriptaddress,showpacs,preprintnumbers,amsmath,amssymb]{revtex4}
\usepackage{bm}% bold math
\usepackage{bbm}
\usepackage{amssymb}
\usepackage{amsfonts}
\usepackage{epsfig,graphicx}
\usepackage{amstext}
\usepackage{amsmath}
\usepackage{graphicx}
\usepackage{times}
\usepackage{dcolumn}% Align table columns on decimal point

\begin{document}
%%%%%%%%%%%%%%%%%%%%%%%%%%%%%%%%%%%%%%%%%%%%%%%%%%%%%%%%%%%%%%%%%%%%%%

%TCIDATA{OutputFilter=Latex.dll}
%TCIDATA{Version=5.00.0.2552}
%TCIDATA{<META NAME="SaveForMode" CONTENT="1">}
%TCIDATA{LastRevised=Wednesday, June 22, 2005 16:21:09}
%TCIDATA{<META NAME="GraphicsSave" CONTENT="32">}

\title{Maximum coherence in the optimal basis}

\author{Ming-Liang Hu}
\email{mingliang0301@163.com}
\affiliation{Institute of Physics, Chinese Academy of Sciences, Beijing 100190, China}
\affiliation{School of Science, Xi'an University of Posts and Telecommunications, Xi'an 710121, China}
\author{Shu-Qian Shen}
\affiliation{College of Science, China University of Petroleum, Qingdao 266580, China}
\author{Heng Fan}
%\email{hfan@iphy.ac.cn}
\affiliation{Institute of Physics, Chinese Academy of Sciences, Beijing 100190, China}
\affiliation{School of Physical Sciences, University of Chinese Academy of Sciences, Beijing 100190, China}
\affiliation{Collaborative Innovation Center of Quantum Matter, Beijing 100190, China}

\begin{abstract}
The resource theoretic measure of quantum coherence is basis
dependent, and the amount of coherence contained in a state is
different in different bases. We obtained analytical solutions for
the maximum coherence by optimizing the reference basis and
highlighted the essential role of the mutually unbiased bases (MUBs)
on attaining the maximum coherence. Apart from the relative entropy
of coherence, we showed that the MUBs are optimal for the robustness
of coherence, the coherence weight, and the modified skew
information measure of coherence for any state. Moreover, the MUBs
are optimal for all the faithful coherence measures if the state is
pure or is of the single qubit. We also highlighted an upper bound
for the $l_1$ norm of coherence and compared it with the other
bounds as well as the maximum one attainable by optimizing the
reference basis.
\end{abstract}


\maketitle

\section{Introduction} \label{sec:1}
Quantum coherence is a basic feature of quantum states that is more
fundamental than the other quantum features such as Bell
nonlocality, entanglement, and quantum discord \cite{Adesso,Hu}. It
also plays an essential role in the study of the reference frames
\cite{fram1,fram2,fram3}, quantum thermodynamics \cite{ther1,ther2,
ther3,ther4,ther5}, and biological systems \cite{bio1,bio2}.
Meanwhile, coherence was recognized to be a precious resource for
various quantum information processing tasks that cannot be
accomplished in a classical way \cite{meas6,asym1,qsm,DJ,qdcohe}.
Despite these importance, a physically meaningful and mathematically
rigorous framework for quantifying coherence was formulated only
recently \cite{coher}. In this framework, the coherence measures
were defined in the similar vein to those of the quantum correlation
measures \cite{Adesso}, with the free states, the free operations,
and the unit resource states being identified explicitly as the
incoherent states, the incoherent operations (IOs), and the
maximally coherent states, respectively \cite{coher}.

Based on the above framework, there are various coherence measures
being proposed until now. Besides the relative entropy of coherence
and $l_1$ norm of coherence \cite{coher}, other faithful measures
include the entanglement-based coherence measure \cite{meas1}, the
robustness of coherence (RoC) \cite{meas6,asym1} and the coherence
weight \cite{cowe}, the modified skew information measure of
coherence \cite{co-ski2}, the two convex roof measures of coherence
which are called intrinsic randomness of coherence \cite{meas4}
(also known as the coherence of formation \cite{dist2}) and
coherence concurrence \cite{measjpa}, the discordlike bipartite
coherence \cite{Guoy}, and an operational coherence measure defined
based on the max-relative entropy \cite{new1}. Moreover, there are
other measures that obey partial of the conditions suggested in Ref.
\cite{coher}, e.g., the skew information measure of coherence, which
is also a well-defined measure of asymmetry \cite{co-ski1}, the
Tsallis relative entropy measure of coherence \cite{meas5}, and the
trace norm of coherence \cite{meas2,meas7}. By identifying the free
operations to be the genuinely incoherent operations which give
$\Phi_{\mathrm{GIO}}(\delta)=\delta$ for all the incoherent states
$\delta\in \mathcal{I}$, the genuine quantum coherence was also
introduced recently \cite{gqc}.

As the coherence measures are basis dependent, it is natural to
wonder in which basis a state attains its maximum amount of
coherence. Or equivalently, what is the optimal unitary that
transforms a state to another state that possesses the maximal
amount of coherence for a fixed reference basis. Here, by saying a
state has the maximal coherence, we mean that its coherence cannot
be enhanced by any unitary transformation, but it is not necessary
to be maximally coherent \cite{coher}. In fact, for the relative
entropy and squared $l_2$ norm of coherence, it has already been
shown that the optimal bases are the mutually unbiased bases (MUBs),
i.e., the bases mutually unbiased to the eigenbasis of the
considered state \cite{maxcoh}. A similar problem has also been
discussed in Ref. \cite{coh-pur}, in which the authors defined the
coherence measures by identifying the free operations as the
maximally incoherent operations (MIOs), namely, $\Phi_{\mathrm{MIO}}
(\delta)\in \mathcal{I}$, $\forall \delta\in \mathcal{I}$, and
showed that the MUBs are optimal for any MIO monotone of coherence.
Moreover, lower bounds for the relative entropy of coherence and the
geometric coherence averaged over a set of MUBs were obtained
\cite{new2}. Despite these progresses, it is noteworthy that there
are non-MIO coherence monotones (e.g., the $l_1$ norm of coherence
\cite{Bu} and the coherence of formation \cite{creat3}) and
coherence measures that have not been proved to be a MIO monotone or
not. Thus, it is worthwhile to identify the optimal bases and the
corresponding maximally attainable coherence of them.

\section{Technical preliminaries} \label{sec:2}
The identification of an optimal basis for attaining the maximum
coherence is equivalent to identifying an optimal unitary for which
the transformed state has the maximum coherence in a fixed basis. As
any density operator $\rho$ can always be diagonalized in the
reference basis spanned by its eigenvectors, that is, by denoting
$V=(|\psi_0\rangle, |\psi_1\rangle, \ldots, |\psi_{d-1} \rangle)$,
with $|\psi_i\rangle$ and $\lambda_i$ denoting, respectively, the
eigenvectors and eigenvalues of $\rho$, we always have $V^\dag \rho
V=\Lambda$, with
%%%%%%%%%%%%%%%%%%%%%%%%%%%
\begin{equation}\label{eq2-1}
 \Lambda=\mathrm{diag} \{\lambda_0, \lambda_1, \ldots, \lambda_{d-1}\},
\end{equation}
%%%%%%%%%%%%%%%%%%%%%%%%%%%
and we denote by $\tilde{U}=\tilde{U}_{\Lambda}V^\dag$ the optimal
unitary for a general state $\rho$, where $\tilde{U}_{\Lambda}$ is
that for the diagonalized state $\Lambda$.

For the Hilbert space $\mathcal {H}$ of arbitrary dimension $d$, one
of the MUB is given by $\{|\phi_m^d\rangle\}$, with
%%%%%%%%%%%%%%%%%%%%%%%%%%%
\begin{equation}\label{eq2-2}
 |\phi_m^d\rangle= \frac{1}{\sqrt{d}}\sum_{n=0}^{d-1}
                 e^{i\frac{2\pi}{d}mn}|n\rangle,
\end{equation}
%%%%%%%%%%%%%%%%%%%%%%%%%%%
where $i$ is the imaginary unit, and $m=0,\ldots, d-1$. Moreover,
for the case of $d$ being a prime, all the $d+1$ MUBs can be
obtained. Apart from $\{|\phi_m^d\rangle\}$ and $\{|\phi_m^0\rangle
= \sum_{n=0}^{d-1}\delta_{mn} |n\rangle\}$, the remaining
$\{|\phi_m^l\rangle\}$ for $l\in\{1,\dots,d-1\}$ are
\cite{MUB1,MUB2}
%%%%%%%%%%%%%%%%%%%%%%%%%%%
\begin{equation}\label{eq2-3}
 |\phi_m^l\rangle= \frac{1}{\sqrt{d}}\sum_{n=0}^{d-1}
                   e^{i\frac{2\pi}{d}l(m+n)^2}|n\rangle.
\end{equation}
%%%%%%%%%%%%%%%%%%%%%%%%%%%

Then if we define the unitary operator
%%%%%%%%%%%%%%%%%%%%%%%%%%%
\begin{equation}\label{eq2-4}
 U_\mathrm{mub}= \sum_{m=0}^{d-1}|\phi_m^k\rangle\langle m|,
\end{equation}
%%%%%%%%%%%%%%%%%%%%%%%%%%%
with $k\in\{1,\dots,d\}$, the transformed state of the unitary
operation $U_\mathrm{mub}$ can be obtained as
%%%%%%%%%%%%%%%%%%%%%%%%%%%
\begin{equation}\label{eq2-5}
 \begin{aligned}
  \tilde{\Lambda}= U_\mathrm{mub}\Lambda U_\mathrm{mub}^\dagger
                 = \sum_{m=0}^{d-1}\lambda_m
                   |\phi_m^k\rangle\langle\phi_m^k|,
 \end{aligned}
\end{equation}
%%%%%%%%%%%%%%%%%%%%%%%%%%%
which is featured for the equal diagonal entries, and it is also
called the contradiagonal density matrix in Ref. \cite{cdbas}.

In general, one can consider the complex Hadamard matrix (CHM)
described by \cite{CHM1,CHM2,CHM3}
%%%%%%%%%%%%%%%%%%%%%%%%%%%
\begin{equation}\label{eq2-6}
 HH^\dagger=d\mathbb{I}, \;
 |H_{ij}|=1~(\forall i,j=0,\dots,d-1),
\end{equation}
%%%%%%%%%%%%%%%%%%%%%%%%%%%
and for convenience of later presentation, we denote by
$H_d=H/\sqrt{d}$ the rescaled CHM.

For any state eigenbasis $\{|\psi_i\rangle\}$, one can see from the
above equations that the basis $\{U_\mathrm{mub}|\psi_i\rangle\}$,
or more general, the basis $\{H_d|\psi_i\rangle\}$, is mutually
unbiased to $\{|\psi_i\rangle\}$. Due to this reason, when referring
to MUBs in the following, we mean those bases mutually unbiased to
$\{|\psi_i\rangle\}$.

\section{Maximum coherence in the MUBs} \label{sec:3}
Based on the above preliminaries, we begin to identify the optimal
basis for the maximum attainable coherence. There are two related
concepts that need to be clarified. The first one is the maximally
coherent state, which was defined to be a measure-independent state
that can serve as a resource for generating all the other states in
the same Hilbert space $\mathcal {H}$ by merely the IOs
\cite{coher}, and the second one is the maximal-coherence-value
states (MCVSs), which have the maximal value of coherence
\cite{peng}. A general state in the optimal basis may not always be
a maximally coherent state or a MCVS.

For the faithful measures of coherence, all the pure states have the
maximum coherence in the MUBs, as for this case they belong to the
set of MCVSs \cite{peng}. For the relative entropy of coherence,
which is a MIO monotone, the MUBs are also optimal, and the
corresponding maximum is given by $\log_2 d-S(\rho)$, where
$S(\rho)=-\mathrm{tr}(\rho \log_2 \rho)$ denotes the von Neumann
entropy \cite{maxcoh,maxrel,coh-pur}. For the other coherence
measures, we report our results in the following text.

We first consider the coherence measured by RoC \cite{meas6}. It was
introduced based on the consideration that the mixture of $\rho$
with another state $\tau$ may be coherent or incoherent, and the
minimal mixing required to destroy completely the coherence in
$\rho$ is defined as the RoC. To be explicit, it is given by
%%%%%%%%%%%%%%%%%%%%%%%%%%%
\begin{equation}\label{eq3-1}
 C_R(\rho)= \min_{\tau\in\mathcal{D}(\mathbb{C}^d)}\left\{s\geq 0 \bigg|
            \frac{\rho+s\tau}{1+s}=: \delta\in\mathcal{I}\right\},
\end{equation}
%%%%%%%%%%%%%%%%%%%%%%%%%%%
where $\mathcal{D}(\mathbb{C}^d)$ is the convex set of density
operators on $\mathcal{H}$, and $\mathcal{I}$ is the set of
incoherent states.

Starting from the above formula, we present our first result via the
following theorem (see Appendix \ref{sec:6} for its proof).

\emph{Theorem 1.} For state $\rho$ of dimension $d$, the maximum RoC
attainable by optimizing the reference basis is
%%%%%%%%%%%%%%%%%%%%%%%%%%%
\begin{equation}\label{eq-therem1}
 C_R^{\max}(\rho)=d\lambda_{\max} -1,
\end{equation}
%%%%%%%%%%%%%%%%%%%%%%%%%%%
where $\lambda_{\max}$ is the largest eigenvalue of $\rho$. The
bases mutually unbiased with the state eigenbasis are the optimal
bases, and $\tilde{U}= H_d V^\dag$ are the optimal unitaries.

While the RoC was defined as the minimal mixing required for
destroying the coherence in a state \cite{meas6}, a similar measure
called coherence weight was proposed recently \cite{cowe}. It reads
%%%%%%%%%%%%%%%%%%%%%%%%%%% \sum\limits_{A\atop B}
\begin{equation}\label{eq4-1}
 C_w(\rho)= \min_{\delta\in \mathcal {I}, \tau\in\mathcal{D}(\mathbb{C}^d)}
            \{s\geq 0| \rho= (1-s)\delta+s\tau\},
\end{equation}
%%%%%%%%%%%%%%%%%%%%%%%%%%%
which corresponds to the minimal weight factor $s$ of coherent state
$\tau$ consumed for preparing $\rho$ on average. It obeys the four
conditions for a faithful measure of coherence \cite{cowe}. Now, we
show that the coherence weight also attains its maximum in the MUBs
(see Appendix \ref{sec:7} for its proof).

\emph{Theorem 2.} For state $\rho$ of dimension $d$, the maximum
coherence weight attainable by optimizing the basis is
%%%%%%%%%%%%%%%%%%%%%%%%%%%
\begin{equation}\label{eq-therem2}
 C_w^{\max}(\rho)=1-d\lambda_{\min},
\end{equation}
%%%%%%%%%%%%%%%%%%%%%%%%%%%
where $\lambda_{\min}$ is the smallest eigenvalue of $\rho$. The
bases mutually unbiased with the state eigenbasis are the optimal
bases, and $\tilde{U}=H_d V^\dag$ are the optimal unitaries.

Third, we consider the Wigner-Yanase (WY) skew information measure
of coherence. The primary measure defined using this quantity was
given by \cite{co-ski1}
%%%%%%%%%%%%%%%%%%%%%%%%%%%
\begin{equation}\label{eq5-1}
 \mathsf{I}(\rho,K)= -\frac{1}{2}\mathrm{tr} \{[\sqrt{\rho},K]^2\},
\end{equation}
%%%%%%%%%%%%%%%%%%%%%%%%%%%
with $K$ being an observable. But it was soon showed to violate the
conditions for a faithful coherence measure \cite{Dubai}. To avoid
this problem, a modified version of coherence measure which also
uses the skew information was proposed \cite{co-ski2}. Compared with
the original definition, it uses $|k\rangle\langle k|$ instead of
$K$ as the observable and was defined as the summation of
$\mathsf{I}(\rho, |k\rangle\langle k|)$ over the basis vectors
$\{|k\rangle\}$, i.e.,
%%%%%%%%%%%%%%%%%%%%%%%%%%%
\begin{equation}\label{eq5-2}
 C_{sk}(\rho)=\sum_k \mathsf{I}(\rho,|k\rangle\langle k|),
\end{equation}
%%%%%%%%%%%%%%%%%%%%%%%%%%%
while for the single-qubit state, it is qualitatively equivalent to
the coherence measure $\mathsf{I}(\rho,K)$ given in Ref.
\cite{co-ski1}.

For this measure of coherence, we have the following theorem (see
Appendix \ref{sec:8} for its proof).

\emph{Theorem 3.} For state $\rho$ of dimension $d$, the maximum
modified skew information measure of coherence attainable by
optimizing the basis is
%%%%%%%%%%%%%%%%%%%%%%%%%%%
\begin{equation}\label{eq-therem3}
 C_{sk}^{\max}(\rho)= 1-\frac{1}{d}\left(\sum_i \sqrt{\lambda_i}\right)^2,
\end{equation}
%%%%%%%%%%%%%%%%%%%%%%%%%%%
where the bases mutually unbiased to the state eigenbasis are the
optimal bases, and $\tilde{U}=H_d V^\dag$ are the optimal unitaries.

It is noteworthy that $C_{sk}^{\max}(\rho)$ equals the total
coherence $C_I(\rho)$ presented in Ref. \cite{maxrel}.

A major concern of any resource theory is how to manipulate the
associated resource states. The above three theorems highlight the
role of the MUBs (or equivalently, the unitaries given by the
rescaled CHMs) on attaining the maximum coherence measured by the
RoC, the coherence weight, and the WY skew information. This is of
practical significance as the amount of coherence inherent within a
state determines its capacity for quantum information processing
\cite{Hu}. For example, in a quantum metrology task \cite{meas6},
$C_R(\rho)$ quantifies the maximum advantage achievable by using
coherent probe $\rho$ as opposed to any incoherent probe $\delta$
\cite{meas6}.

The rescaled CHMs are optimal for attaining the maximum RoC and
coherence weight can also be understood by combining Theorems 1 and
2 with the fact that $M$ (see Appendix \ref{sec:6}) is the
coherence-value-preserving operation (CVPO) which conserves the
coherence values of all states, or equivalently, any set of MCVSs is
unchanged under CVPO \cite{peng}. Consequently, provided that the
basis $\{|\phi_m^d\rangle\}$ is optimal, all the bases given by the
equivalent class of CHMs are optimal. That is, by using $H_d V^\dag$
any state can be transformed unitarily into another state that has
the maximum attainable coherence. We noted that the RoC and the
fidelity-based coherence measure are also MIO monotones as they are
monotones of the special R\'{e}nyi relative entropies
\cite{Gour,Zhuhj}, hence the finding that $H_d V^\dag$ is optimal
can also be understood from \cite{coh-pur}. But deriving the maximum
attainable fidelity-based coherence is still a hard task for general
states.

It is also noteworthy that Theorem 2 implies that the coherence
weight for all the rank deficient states take the maximum 1 in the
MUBs. Or equivalently, all the rank deficient states can be
transformed unitarily into the MCVSs in the sense that
$C_w^{\max}(\rho)=1$ for them. This is a property of the coherence
weight that differs it from the other coherence measures, as the
latter are maximal only for the MCVSs given in Eq. (2) of Ref.
\cite{peng}. This also implies that the set of MCVSs may be
different for different coherence measures.

Moreover, for the single qubit case, the MUBs are optimal for
arbitrary coherence measure formulated in the framework of Baumgratz
\textit{et al.} \cite{coher}. This can be proved by noting that the
$l_1$ norm of coherence for a single qubit is $C_{l_1}(\rho)=
(r_1^2+r_2^2)^{1/2}\equiv r_{12}$, where $r_i=\mathrm{tr} (\rho
\sigma_i)$, and $\sigma_{1,2,3}$ are the Pauli operators. It implies
that $r'_{12}$ for $\Phi_{\mathrm{IO}} (\rho)$ cannot be larger than
$r_{12}$ for $\rho$ (see also \cite{qubit}). Meanwhile, by its
definition we know that any IO monotone of coherence $C$ should not
be increased by the IO. Then if $r_{12}\geq r'_{12}$, we always have
%%%%%%%%%%%%%%%%%%%%%%%%%%%
\begin{equation}
 C(r_{12})\geq C(r'_{12}).
\end{equation}
%%%%%%%%%%%%%%%%%%%%%%%%%%%
On the other hand, $r_{12}$ is maximal in the MUBs for any single
qubit state $\rho$, thus any IO monotone $C(\rho)$ attains its
maximum value in the MUBs.

One can check the above result via all the known IO monotones of
coherence such as the $l_1$ norm and relative entropy of coherence
\cite{coher}, the fidelity-based measure of coherence \cite{meas1},
the intrinsic randomness of coherence \cite{meas4}, the coherence of
formation \cite{dist2}, and the coherence concurrence
\cite{measjpa}.

\section{$l_1$ norm of coherence} \label{sec:4}
The $l_1$ norm of coherence is $C_{l_1}(\rho) = \sum_{i\neq
j}|\langle i|\rho|j\rangle|$ in the basis $\{|i\rangle\}$
\cite{coher}, which is favored for the compact analytical solution.
It was formulated in the framework of Baumgratz \textit{et al.}
\cite{coher} and does not obey the additional conditions (C5) and
(C6) presented in Ref. \cite{Adesso}. It has been shown that
$C_{l_1}(U\rho U^\dag)$ is upper bounded by \cite{rqc}
%%%%%%%%%%%%%%%%%%%%%%%%%%%
\begin{equation}\label{eq-bd}
 \mathcal{B}_d= \sqrt{(d^2-d)/2}|\vec{x}|,
\end{equation}
%%%%%%%%%%%%%%%%%%%%%%%%%%%
where $\vec{x}=(x_1,x_2,\ldots,x_{d^2-1})$ is the Bloch vector for
$\rho$, with $x_i=\mathrm{tr} (\rho X_i)$, and $\{X_i\}$ denoting
generators of the Lie algebra $\mathrm{SU}(d)$ \cite{sud1,sud2}.

The bound $\mathcal{B}_d$ is intimately related to the quantumness
captured by noncommutativity of the algebra of observables and the
complementarity of quantum coherence under MUBs \cite{rqc}. It also
has immediate consequence on quantum entanglement. This is because
any bipartite entangled state must have a linear purity $\mathrm{tr}
\rho^2\geq 1/(d-1)$ \cite{lipur}, which is equivalent to
$\mathcal{B}_d \geq 1$. Hence, any bipartite entangled state must
have the coherence larger than a critical value. Moreover, as
$\mathrm{tr}\rho^2=1/d+|\vec{x}|^2/2$, and all the unitary
operations $U$ do not change the mixedness $M_l(\rho)=
d(1-\mathrm{tr} \rho^2)/(d-1)$ of a state, one can give an
alternative proof of the complementarity relation obtained in
\cite{comple2} by using the bound $\mathcal{B}_d$.

For any pure state and single-qubit state, it has been shown that
the bound $\mathcal{B}_d$ can always be reached, and the MUBs are
the optimal reference bases. For general qutrit states, one can also
show that the MUBs are optimal. This is because for
$\tilde{\Lambda}$ of Eq. \eqref{eq2-5} with $d=3$, we always have $
C_{l_1}(\tilde{\Lambda})= \mathcal{B}_3$. As the bound
$\mathcal{B}_3$ cannot be exceed by any unitary equivalent states of
$\tilde{\Lambda}$, the unitaries $\tilde{U}=H_d V^\dag$ are optimal
for any qutrit state.

For $d\geq 4$, $H_d V^\dag$ is not optimal, in general
\cite{maxcoh}. But one can see that for the family of
\textit{physically allowed} (i.e., $\rho_{\mathrm{mcms}}\geq 0$)
states
%%%%%%%%%%%%%%%%%%%%%%%%%%%
\begin{equation} \label{eq6-2}
 \rho_{\mathrm{mcms}}= \frac{1}{d}\mathbb{I}_d+\frac{R}{2}\sum\limits_{i,j=0 \atop i<j}^{d-1}
                       (e^{i\phi_{ij}}|i\rangle\langle j| + e^{-i\phi_{ij}}|j\rangle\langle i|),
\end{equation}
%%%%%%%%%%%%%%%%%%%%%%%%%%%
with $R=2\sqrt{1-M_f}/d$, and $M_f$ the mixedness of
$\rho_{\mathrm{mcms}}$, the bound $\mathcal{B}_d$ is reached. Here,
the phases $\{\phi_{ij}\}$ should satisfy the positive semidefinite
constraints of $\rho_{\mathrm{mcms}}$. It covers the full set of
maximally coherent mixed states (MCMSs). In \cite{comple2}, the
authors stated that $\{\phi_{ij}\}$ can be removed by the incoherent
unitary $U= \sum_{n=0}^{d-1}e^{-i\gamma_n} |n\rangle \langle n|$,
with $\phi_{ij}=\gamma_i-\gamma_j$. But apart from some very special
cases, one can check directly that there does not exist such
$\{\gamma_i\}$ in general. Consequently, $\rho_m$ given in
\cite{comple2} is only a subset of the MCMSs.

%%%%%%%%%%%%%%%%%%%%%%%%%%%%%%%
\begin{figure}
\centering
\resizebox{0.45 \textwidth}{!}{%
\includegraphics{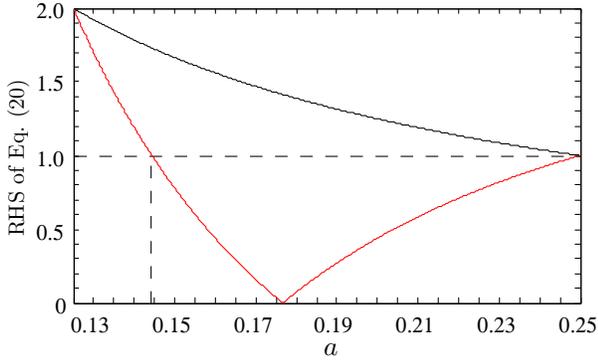}}
% If not, use\vspace{5cm}
% Give the correct figure height in cm
\caption{The absolute value of the right-hand side of Eq.
\eqref{eq6-6} versus $a$, with the sign of $\sin\theta_1$ and
$\sin\theta_3$ being the same (the black solid line) and different
(the red solid line).} \label{fig1}
% Give an unique label
\end{figure}
%%%%%%%%%%%%%%%%%%%%%%%%%%%%%%%

Now, the question is whether there exists an optimal basis (or an
optimal unitary $\tilde{U}$) such that the bound $\mathcal {B}_d$ is
saturated for $d\geq 4$. We show through a counterexample that this
is not always true. To this end, we consider the case of a rank-2
state $\Lambda=\{\lambda_0,\lambda_1,0,0\}$. Then if there exists
such an $\tilde{U}$, it must transform $\Lambda$ to
$\rho_{\mathrm{mcms}}$ of Eq. \eqref{eq6-2}. As the unitary
transformation does not change the rank of a state, all of the
third-order minors of $\rho_{\mathrm{mcms}}$ must be zero
\cite{matrix}. But this cannot always be satisfied. This is because
for $d=4$, $\rho_{\mathrm{mcms}}$ is incoherent unitary equivalent
to
%%%%%%%%%%%%%%%%%%%%%%%%%%%
\begin{equation}\label{eq6-3}
 \Lambda_U =
   \left(\begin{array}{cccc}
      \frac{1}{4}       & a                 & a e^{i\theta_1}  & a e^{i\theta_2} \\
      a                 & \frac{1}{4}       & a                & a e^{i\theta_3} \\
      a e^{-i\theta_1}  & a                 & \frac{1}{4}      & a \\
      a e^{-i\theta_2}  & a e^{-i\theta_3}  & a                & \frac{1}{4}\\
  \end{array}\right),
\end{equation}
%%%%%%%%%%%%%%%%%%%%%%%%%%%
where $a=R/2\in[1/8,1/4]$. The third-order leading principal minor
can be calculated as
%%%%%%%%%%%%%%%%%%%%%%%%%%%
\begin{equation} \label{eq6-4}
 D_3= \frac{1}{64}-\frac{3}{4}a^2+2a^3\cos\theta_1,
\end{equation}
%%%%%%%%%%%%%%%%%%%%%%%%%%%
from which we obtain $\cos\theta_1=3/8a-1/128a^3$. Similarly, we
have $\cos\theta_3=\cos\theta_1$ (but $\sin\theta_3$ and
$\sin\theta_1$ may be different). Further, the minor of the
principal submatrix formed by removing from $\Lambda_U$ its first
row and last column is
%%%%%%%%%%%%%%%%%%%%%%%%%%%
\begin{equation}\label{eq6-5}
\begin{aligned}
 \Delta_3= & a^3[1+e^{-i(\theta_1+\theta_3)}]-\frac{1}{4}a^2(e^{-i\theta_1}+e^{-i\theta_3}) \\
           & +(\frac{1}{16}a-a^3)e^{-i\theta_2},
\end{aligned}
\end{equation}
%%%%%%%%%%%%%%%%%%%%%%%%%%%
and $\Delta_3=0$ requires
%%%%%%%%%%%%%%%%%%%%%%%%%%%
\begin{equation}\label{eq6-6}
\begin{aligned}
 e^{-i\theta_2}= \frac{16a^2[1+e^{-i(\theta_1+\theta_3)}]-4a(e^{-i\theta_1}
                 +e^{-i\theta_3})]}{1-16a^2}.
\end{aligned}
\end{equation}
%%%%%%%%%%%%%%%%%%%%%%%%%%%
But from Fig. \ref{fig1} one can note that except the cases
$a=1/4\sqrt{3}$ and $1/4$, there are no solutions for $\Delta_3=0$.
This implies that the rank-2 state $\Lambda$ may not be transformed
to a MCMS by any unitary, thus the bound $\mathcal{B}_d$ cannot
always be reached.

%%%%%%%%%%%%%%%%%%%%%%%%%%%%%%%
\begin{figure}
\centering
\resizebox{0.45 \textwidth}{!}{%
\includegraphics{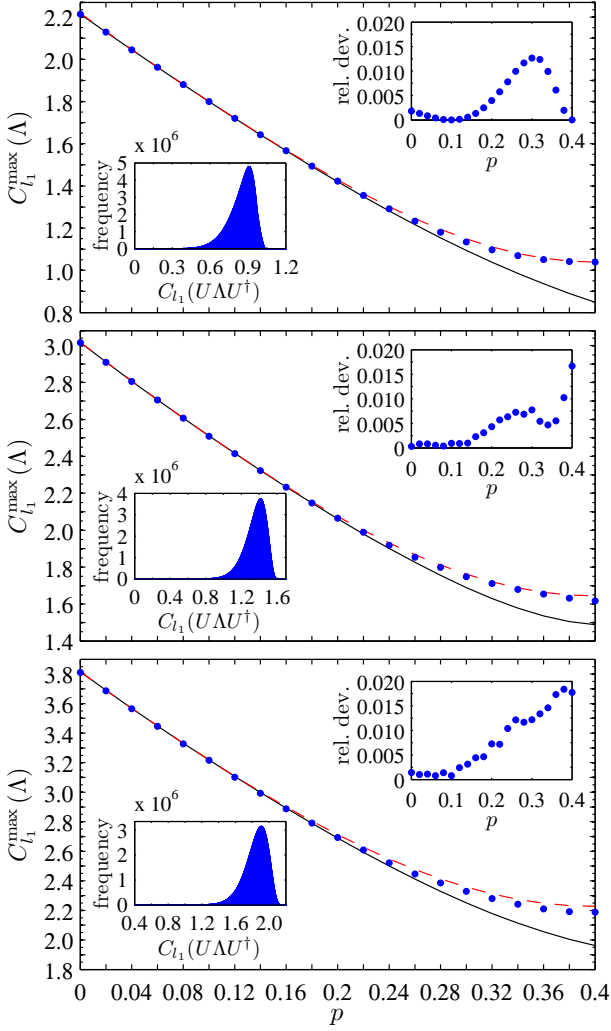}}
% If not, use\vspace{5cm}
% Give the correct figure height in cm
\caption{Maximum $l_1$ norm of coherence $C_{l_1}^{\max}(\Lambda)$
for $d=4$, 5, and 6 (from top to bottom), where the black solid
lines (red dashed lines) correspond to $\mathcal{O}_d$
($\mathcal{B}_d$), and every blue solid circle was the numerical
result obtained by $10^9$ equally distributed unitaries generated
according to the Haar measure. Moreover, the insets at the top-right
corner are the relative deviations of $C_{l_1}^{\max} (\Lambda)$
from $\mathcal{B}_d$, and the insets at the bottom-left corner are
the distribution of $C_{l_1}(U\Lambda U^\dag)$ for $p=0.4$.}
\label{fig2}
% Give an unique label
\end{figure}
%%%%%%%%%%%%%%%%%%%%%%%%%%%%%%%

Although $\mathcal{B}_d$ may not be reached in general, our
numerical results present strong evidence that the difference
between it and $C_{l_1}^{\max}(\Lambda)$ may be small. For the
rank-2 state $\Lambda$ , we have performed calculations by choosing
$\lambda_0=0.05k$ ($k\in \mathbb{Z}$), and obtained the maximum
$C_{l_1}^{\max}(\Lambda)$ via $10^9$ equally distributed unitaries
$U$ generated according to the Haar measure \cite{Haar,Haar2}. The
results showed that the relative deviations of $C_{l_1}^{\max}$ from
$\mathcal{B}_d$, i.e.,
%%%%%%%%%%%%%%%%%%%%%%%%%%%
\begin{equation}
 \frac{\mathcal{B}_d-C_{l_1}^{\max}}{\mathcal{B}_d},
\end{equation}
%%%%%%%%%%%%%%%%%%%%%%%%%%%
are between $1.392836\times 10^{-4}$ and 0.012743. It is also
noteworthy that this deviation is upper bounded by $1 -
\mathcal{O}_d/\mathcal{B}_d$.

In Fig. \ref{fig2}, we further showed the exemplified plots for
states $\Lambda= \mathrm{diag} \{0.1,0.1,p, 0.8-p\}$, $\mathrm{diag}
\{0.04,0.06,0.1, p,0.8-p\}$, and $\mathrm{diag} \{0.02,
0.04,0.06,0.08, p,0.8-p\}$, respectively. For comparison, we plotted
in the same figure also the bound $\mathcal{O}_d$ given in Ref.
\cite{maxcoh}, i.e.,
%%%%%%%%%%%%%%%%%%%%%%%%%%%
\begin{equation}\label{eq6-7}
 \mathcal{O}_d= \sum_{n=1}^{d-1}\sqrt{\sum_{i=0}^{d-1}\lambda_i^2 +
                \sum_{k\neq l}^{d-1}\lambda_k \lambda_l
                \cos\left[\frac{2\pi n}{d}(k-l)\right]}.
\end{equation}
%%%%%%%%%%%%%%%%%%%%%%%%%%%

From these plots one can see that in the regions of relative small
$p$, the two bounds $\mathcal{O}_d$ and $\mathcal{B}_d$ give nearly
the same estimation for the maximally achievable $l_1$ norm of
coherence. But with the increase of $p$, $C_{l_1}^{\max}$ turns to
violate those of $\mathcal{O}_d$, and are still very close to their
upper bound $\mathcal{B}_d$. In the top-right insets of Fig.
\ref{fig2}, we also plotted the relative deviations of
$C_{l_1}^{\max}$ from $\mathcal{B}_d$. The results revealed that
they are between $1.893424\times 10^{-4}$ and 0.018391 for the
considered data points. It is also expectable that with the
increasing dimension $d$, one should perform more and more runs of
simulation to reduce the relative deviation due to the structure of
the uniformly distributed Haar measure \cite{Haar,Haar2}. This
explains why the relative deviations are increased with $d$ in
general when the same runs of simulation are performed.

We have also calculated the distribution of $C_{l_1}(U\Lambda
U^\dag)$ for $p=0.4$ with $10^9$ Haar distributed unitaries. The
results were displayed in the insets at the bottom-left corner of
Fig. \ref{fig2}. For $d=4$, 5, and 6, the probabilities for
$C_{l_1}(U\Lambda U^\dag)> \mathcal{O}_d$ is about $57.65\%$,
$10.86\%$, and $20.30\%$. This confirms again that $\mathcal{O}_d$
can be overcame for $d\geq 4$ \cite{maxcoh}. In particular, the peak
value of $C_{l_1}(U\Lambda U^\dag)$ ($\sim 0.908$) is also larger
than $\mathcal{O}_4 \simeq 0.848528$ for $d=4$. Of course, for the
other two cases, the peak values ($\sim 1.415$ and 1.911) turn out
to be smaller than $\mathcal{O}_5 \simeq 1.488135$ and
$\mathcal{O}_6 \simeq 1.961348$.

Finally, by combining Eq. \eqref{eq-therem1} with the results of
Refs. \cite{meas6,asym1}, one can obtain another upper bound for the
maximum $l_1$ norm of coherence, which is given by
%%%%%%%%%%%%%%%%%%%%%%%%%%%
\begin{equation}\label{eq-new1}
 \mathcal{R}_d= (d-1)(d\lambda_{\max}-1),
\end{equation}
%%%%%%%%%%%%%%%%%%%%%%%%%%%
then it is natural to ask whether this bound could give a better
estimation for $C_{l_1}^{\max}$ than $\mathcal{B}_d$. But a direct
calculation shows that this is not the case. This is because for
$\Lambda$ of Eq. \eqref{eq2-1}, we always have $\mathcal{R}_d \geq
\mathcal{B}_d$ (see Appendix \ref{sec:9} for its proof). Thus, the
bound $\mathcal{R}_d$ is not better than $\mathcal{B}_d$.

\section{Summary} \label{sec:5}
In summary, we have studied the maximal amount of quantum coherence
attainable by optimizing the reference basis, or equivalently, by
performing optimal unitary operations on the state in a fixed basis.
For the RoC, the coherence weight, and the modified skew information
measure of coherence, we obtained analytical solutions for their
maximum values, and proved strictly that the optimal bases are the
MUBs, i.e., the bases mutually unbiased with the eigenbasis of the
considered state. Moreover, when considering the pure states and the
single qubit states, the MUBs are optimal for any IO monotone of
coherence. While these highlight role of the MUBs on attaining the
maximum coherence, they fail for the $l_1$ norm of coherence for
states $\rho$ of dimension $d\geq 4$. We emphasized the upper bound
$\mathcal{B}_d$ of $C_{l_1}(\rho)$, and showed that for the rank-2
state it may not always be reached. We also presented strong
evidence that the difference between the maximum attainable $l_1$
norm of coherence and the bound $\mathcal{B}_d$ may be small in most
cases, though a strict proof is still needed.

By combining the present work with Refs. \cite{maxcoh,coh-pur,
maxrel}, it also seems that apart from the $l_1$ norm of coherence,
most of the other well-known faithful coherence measures are maximal
in the MUBs. Then there is an interesting question as to whether the
coherence defined in the MUBs can be dubbed an intrinsic coherent
property of quantum states? At least this can avoid the
basis-dependent perplexity of various coherence measures, as in most
cases we are inclined to scrutinize properties of a system via a
basis-independent quantity.

\section*{ACKNOWLEDGMENTS}
We thank Huangjun Zhu and Karol \.{Z}yczkowski for valuable
discussion and correspondence. This work was supported by NSFC
(Grants No. 11675129 and No. 91536108), MOST (Grants No.
2016YFA0302104 and No. 2016YFA0300600), New Star Project of Science
and Technology of Shaanxi Province (Grant No. 2016KJXX-27), CAS
(Grants No. XDB01010000 and No. XDB21030300), and New Star Team of
XUPT. S.-Q.S. was supported by the Natural Science Foundation of
Shandong Province (Grant No. ZR2016AM23) and the Fundamental
Research Funds for the Central Universities (Grant No. 15CX05062A).

%%%%%%%%%%%%%%%%%%%%%%%%%%%%%%%%%%%%%%%%%%%%%%%%%%%%%%%%%%%%%%%%%%
%%%%%%%%%%%%%%%%%%%%%%%%%%%%%%%%%%%%%%%%%%%%%%%%%%%%%%%%%%%%%%%%%%
\begin{appendix}
\section{Proof of Theorem 1} \label{sec:6}
\setcounter{equation}{0}
\renewcommand{\theequation}{A\arabic{equation}}
\setcounter{figure}{0}
\renewcommand{\thefigure}{A\arabic{figure}}
By virtue of Eq. \eqref{eq3-1} one can see that when the mixture of
$\tilde{\Lambda}$ and $\delta$ is incoherent, the required
$\tilde{\tau}$ should be of the form $\tilde{\tau}= [(1+s)\delta
-\tilde{\Lambda}]/s$. The positive semidefiniteness of it requires
$\langle\vec{x}, \tilde{\tau} \vec{x} \rangle \geq 0$, $\forall~
\vec{x}\neq 0$, where $\langle u,v\rangle=\mathrm{tr}(u^\dagger v)$
is the inner product \cite{matrix}. By choosing
$\vec{x}=(1,1,\ldots, 1)^T$ (the superscript $T$ denotes transpose),
assuming $\lambda_0 \geq \lambda_1\geq\ldots \geq\lambda_{d-1}$, and
further using the relation
%%%%%%%%%%%%%%%%%%
\begin{equation}\label{eqa-1}
 \langle\vec{x}, |\phi_m^d\rangle\langle\phi_m^d|\vec{x}\rangle=\left\{
    \begin{aligned}
     &d    && \mathrm{if}~ m =0,\\
     &0    && \mathrm{if}~ m=1,\dots,d-1,
    \end{aligned} \right.
\end{equation}
%%%%%%%%%%%%%%%%%%
one can derive
%%%%%%%%%%%%%%%%%%%%%%%%%%%
\begin{equation}\label{eqa-2}
 \langle\vec{x},\tilde{\tau}\vec{x}\rangle= \frac{1}{s}(1+s-d\lambda_0).
\end{equation}
%%%%%%%%%%%%%%%%%%%%%%%%%%%

Similarly, for $|\phi_m^l\rangle$ with $l=1,\dots, d-1$, by defining
$\vec{x}'= U_I^\dagger \vec{x}$ with
%%%%%%%%%%%%%%%%%%
\begin{equation}\label{eqa-3}
 U_I=\sum_{n=0}^{d-1} e^{-i\frac{2\pi}{d}l n^2}|n\rangle\langle n|,
\end{equation}
%%%%%%%%%%%%%%%%%%
one can show that the inner product $\langle\vec{x}', \tilde{\tau}
\vec{x}' \rangle$ is the same as Eq. \eqref{eqa-2}, then the
positive semidefiniteness of $\tilde{\tau}$ implies $s\geq
d\lambda_0 -1$. As the RoC is defined to be the minimal weight of
mixing, we have $C_R(\tilde{\Lambda}) \geq d\lambda_0 -1$. This,
together with the bound $C_R(\tilde{\Lambda}) \leq d\lambda_0 -1$
given in \cite{meas6}, yields Eq. \eqref{eq-therem1}.

For the rescaled CHM, as one can always find an incoherent unitary
$M$ such that  the entries in the first column of $M H M^\dagger$
are 1. Then by denoting $\vec{x}'=M^\dagger \vec{x}$ and
$|\varphi_m\rangle$ the $m$th column of $H_d$, one can show via Eq.
\eqref{eq2-6} that $\langle\vec{x}', |\varphi_m \rangle \langle
\varphi_m| \vec{x}'\rangle$ is the same as Eq. \eqref{eqa-2}. Hence
we still have $s\geq d\lambda_0 -1$, which further gives Eq.
\eqref{eq-therem1}. This completes the proof.

\section{Proof of Theorem 2} \label{sec:7}
\setcounter{equation}{0}
\renewcommand{\theequation}{B\arabic{equation}}
First, for every state $\rho$ it holds that
%%%%%%%%%%%%%%%%%%%%%%%%%%%
\begin{equation}\label{eqb-1}
 \rho\geq \lambda_{\min}\mathbb{I}_d=d\lambda_{\min}\frac{\mathbb{I}_d}{d},
\end{equation}
%%%%%%%%%%%%%%%%%%%%%%%%%%%
with $\mathbb{I}_d /d$ being the maximally mixed state that is
obviously incoherent. This, together with the alternative definition
of the coherence weight $C_w(\rho)=\min_{\delta\in \mathcal
{I}}\{s\geq 0|\rho\geq (1-s)\delta\}$ \cite{cowe}, implies
immediately that
%%%%%%%%%%%%%%%%%%%%%%%%%%%
\begin{equation}\label{eqb-2}
 C_w(\rho)\leq 1-d\lambda_{\min}.
\end{equation}
%%%%%%%%%%%%%%%%%%%%%%%%%%%

Second, from Eq. \eqref{eq4-1} we know that for $\tilde{\Lambda}$,
the required $\tilde{\tau}$ is $\tilde{\tau}= [\tilde{\Lambda}
-(1-s)\delta]/s$, thus by assuming $\lambda_0 \leq \lambda_1 \leq
\ldots\leq \lambda_{d-1}$, $\vec{x}=(1,1,\ldots, 1)^T$, and using
Eq. \eqref{eqa-1}, we obtain
%%%%%%%%%%%%%%%%%%%%%%%%%%%
\begin{equation}\label{eqb-3}
 \langle\vec{x}, \tilde{\tau} \vec{x} \rangle =1+\frac{1}{s}(d\lambda_0-1).
\end{equation}
%%%%%%%%%%%%%%%%%%%%%%%%%%%

Then the positive semidefiniteness of $\tilde{\tau}$ implies $s\geq
1-d\lambda_0$. Note that here we denote by $\lambda_0$ the smallest
eigenvalue, hence $C_w(\tilde{\Lambda})\geq 1- d\lambda_{\min}$. By
comparing this with Eq. \eqref{eqb-2}, we arrive at Eq.
\eqref{eq-therem2}. Furthermore, with the help of the same
$\vec{x}'$ for proving Theorem 1, one can show that any general
bases given by the rescaled CHMs are optimal for attaining the
maximum coherence weight. This completes the proof.

\section{Proof of Theorem 3} \label{sec:8}
\setcounter{equation}{0}
\renewcommand{\theequation}{C\arabic{equation}}
By using an equivalent definition of $C_{sk}(\rho)$ \cite{co-ski2}
%%%%%%%%%%%%%%%%%%%%%%%%%%%
\begin{equation}\label{eqc-1}
 C_{sk}(\rho)=1-\sum_k \langle k|\sqrt{\rho} |k\rangle^2,
\end{equation}
%%%%%%%%%%%%%%%%%%%%%%%%%%%
one can show that
%%%%%%%%%%%%%%%%%%%%%%%%%%%
\begin{equation}\label{eqc-2}
\begin{aligned}
 C_{sk}(U\Lambda U^\dag) &= 1-\sum_k  \left(\sum_i\sqrt{\lambda_i}
                 \langle k| \psi_i\rangle\langle \psi_i |k\rangle\right)^2 \\
              &= 1-\sum_k \left( \sum_i \sqrt{\lambda_i} k_i^2 \right)^2 \\
              &\leq 1-\frac{1}{d}\left(\sum_k \sum_i \sqrt{\lambda_i} k_i^2 \right)^2 \\
              &= 1-\frac{1}{d}\left(\sum_i \sqrt{\lambda_i} \right)^2,
\end{aligned}
\end{equation}
%%%%%%%%%%%%%%%%%%%%%%%%%%%
where $|\psi_i\rangle= U|i\rangle$, and the overlap $k_i=|\langle
k|\psi_i\rangle|$. The inequality comes from the fact that the
arithmetic mean of a list of nonnegative real numbers is not larger
than the quadratic mean of the same list, i.e.,
%%%%%%%%%%%%%%%%%%%%%%%%%%%
\begin{equation}\label{eqc-3}
 \frac{1}{d}\sum_i a_i \leq \sqrt{\frac{1}{d}\sum_i a_i^2},
\end{equation}
%%%%%%%%%%%%%%%%%%%%%%%%%%%
and the last equality is due to $\sum_k k_i^2 =1$, $\forall i=0,
\ldots,d-1$.

The equality condition  in Eq. \eqref{eqc-2} holds when $\sum_i
\sqrt{\lambda_i} k_i^2$ are the same for different $k$, namely, when
the basis $\{|\psi_i\rangle\}$ is mutually unbiased to
$\{|k\rangle\}$. Clearly, $H_d$ satisfies this requirement, as it
gives $|\langle k|H_d |i\rangle|= 1/\sqrt{d}$, $\forall
i,k=0,1,\dots,d-1$. This completes the proof.\\

\section{Proof of $\mathcal{R}_d \geq \mathcal{B}_d$} \label{sec:9}
\setcounter{equation}{0}
\renewcommand{\theequation}{D\arabic{equation}}
For state $\Lambda$ of Eq. \eqref{eq2-1}, by assuming $\lambda_0
\geq \lambda_1\geq\ldots \geq\lambda_{d-1}$, we have
%%%%%%%%%%%%%%%%%%%%%%%%%%%
\begin{equation}\label{eqd-1}
 \begin{aligned}
  \mathcal{R}_d^2=&(d-1)^2 \Bigg[(d-1)^2\lambda_0^2+\sum_{j=1}^{d-1}\lambda_j^2 \\
                  &-2(d-1)\lambda_0\sum_{j=1}^{d-1}\lambda_j
                   +2\sum_{i=1<j}^{d-1}\lambda_i\lambda_j \Bigg],
 \end{aligned}
\end{equation}
%%%%%%%%%%%%%%%%%%%%%%%%%%%
and
%%%%%%%%%%%%%%%%%%%%%%%%%%%
\begin{equation}\label{eqd-2}
 \begin{aligned}
  \mathcal{B}_d^2=& (d-1) \Bigg[(d-1) \sum_{j=0}^{d-1}\lambda_j^2
                    -2\lambda_0\sum_{j=1}^{d-1}\lambda_j \\
                  & -2\sum_{i=1<j}^{d-1}\lambda_i\lambda_j \Bigg].
 \end{aligned}
\end{equation}
%%%%%%%%%%%%%%%%%%%%%%%%%%%
Then one can obtain directly that
%%%%%%%%%%%%%%%%%%%%%%%%%%%
\begin{equation}\label{eqd-3}
 \begin{aligned}
  \mathcal{R}_d^2-\mathcal{B}_d^2=& d(d-1) \Bigg[(d-1)(d-2)\lambda_0^2 \\
                                  & -2(d-2)\lambda_0 \sum_{j=1}^{d-1}\lambda_j
                                    +2\sum_{i=1<j}^{d-1}\lambda_i\lambda_j\Bigg] \\
                                 =& d(d-1)\sum_{i=1<j}^{d-1}
                                    (\lambda_0-\lambda_i)(\lambda_0-\lambda_j),
 \end{aligned}
\end{equation}
%%%%%%%%%%%%%%%%%%%%%%%%%%%
which implies $\mathcal{R}_d\geq \mathcal{B}_d$.

\end{appendix}

\newcommand{\PRL}{Phys. Rev. Lett. }
\newcommand{\RMP}{Rev. Mod. Phys. }
\newcommand{\PRA}{Phys. Rev. A }
\newcommand{\PRB}{Phys. Rev. B }
\newcommand{\PRE}{Phys. Rev. E }
\newcommand{\PRX}{Phys. Rev. X }
\newcommand{\NJP}{New J. Phys. }
\newcommand{\JPA}{J. Phys. A }
\newcommand{\JPB}{J. Phys. B }
\newcommand{\PLA}{Phys. Lett. A }
\newcommand{\NP}{Nat. Phys. }
\newcommand{\NC}{Nat. Commun. }
\newcommand{\SR}{Sci. Rep. }
\newcommand{\EPJD}{Eur. Phys. J. D }
\newcommand{\QIP}{Quantum Inf. Process. }
\newcommand{\QIC}{Quantum Inf. Comput. }
\newcommand{\AoP}{Ann. Phys. }
\newcommand{\PR}{Phys. Rep. }
%
% BibTeX users please use
% \bibliographystyle{}
% \bibliography{}

\begin{thebibliography}{50}

%%%%%%%%%%%%%%%%%%%%%%%%%%%%%%%
\bibitem{Adesso} A. Streltsov, G. Adesso, and M. B. Plenio, arXiv:1609.02439.
\bibitem{Hu} M. L. Hu, X. Hu, J. C. Wang, Y. Peng, Y. R. Zhang, and H. Fan, arXiv:1703.01852.
\bibitem{fram1} S. D. Bartlett, T. Rudolph, and R. W. Spekkens, \RMP {\bf 79}, 555 (2007).
\bibitem{fram2} I. Marvian and R. W. Spekkens, \NJP {\bf 15}, 033001 (2013).
\bibitem{fram3} I. Marvian and R. W. Spekkens, \PRA {\bf 90}, 062110 (2014).
\bibitem{ther1} J. {\AA}berg, \PRL {\bf 113}, 150402 (2014).
\bibitem{ther2} M. Lostaglio, D. Jennings, and T. Rudolph, \NC {\bf 6}, 6383 (2015).
\bibitem{ther3} V. Narasimhachar and G. Gour, \NC {\bf 6}, 7689 (2015).
\bibitem{ther4} M. Lostaglio, K. Korzekwa, D. Jennings, and T. Rudolph, \PRX {\bf 5}, 021001 (2015).
\bibitem{ther5} P. \'{C}wikli\'{n}ski, M. Studzi\'{n}ski, M. Horodecki, and J. Oppenheim, \PRL {\bf 115}, 210403 (2015).

\bibitem{bio1} N. Lambert, Y.-N. Chen, Y.-C. Cheng, C.-M. Li, G.-Y. Chen, and F. Nori, \NP {\bf 9}, 10 (2013).
\bibitem{bio2} S. Lloyd, J. Phys. Conf. Ser. {\bf 302}, 012037 (2011).
\bibitem{meas6} C. Napoli, T. R. Bromley, M. Cianciaruso, M. Piani, N. Johnston, and G. Adesso, \PRL {\bf 116}, 150502 (2016).
\bibitem{asym1} M. Piani, M. Cianciaruso, T. R. Bromley, C. Napoli, N. Johnston, and G. Adesso, \PRA {\bf 93}, 042107 (2016).
\bibitem{qsm} A. Streltsov, E. Chitambar, S. Rana, M. N. Bera, A. Winter, and M. Lewenstein, \PRL {\bf 116}, 240405 (2016).
\bibitem{DJ} M. Hillery, \PRA {\bf 93}, 012111 (2016).
\bibitem{qdcohe} J. Ma, B. Yadin, D. Girolami, V. Vedral, and M. Gu, \PRL {\bf 116}, 160407 (2016).
\bibitem{coher} T. Baumgratz, M. Cramer, and M. B. Plenio, \PRL {\bf 113}, 140401 (2014).
\bibitem{meas1} A. Streltsov, U. Singh, H. S. Dhar, M. N. Bera, and G. Adesso, \PRL {\bf 115}, 020403 (2015).
\bibitem{cowe} K. Bu, N. Anand, and U. Singh, arXiv:1703.01266.

\bibitem{co-ski2} C. S. Yu, \PRA {\bf 95}, 042337 (2017).
\bibitem{meas4} X. Yuan, H. Zhou, Z. Cao, and X. Ma, \PRA {\bf 92}, 022124 (2015).
\bibitem{dist2} A. Winter and D. Yang, \PRL {\bf 116}, 120404 (2016).
\bibitem{measjpa} X. Qi, T. Gao, and F. Yan, \JPA {\bf 50}, 285301 (2017).
\bibitem{Guoy} Y. Guo and S. Goswami, \PRA {\bf 95}, 062340 (2017).
\bibitem{new1} K. Bu, U. Singh, S. M. Fei, A. K. Pati, and J. Wu, arXiv:1707.08795.
\bibitem{co-ski1} D. Girolami, \PRL {\bf 113}, 170401 (2014).
\bibitem{meas5} A. E. Rastegin, \PRA {\bf 93}, 032136 (2016).
\bibitem{meas7} S. Rana, P. Parashar, and M. Lewenstein, \PRA {\bf 93}, 012110 (2016).
\bibitem{meas2} L. H. Shao, Z. Xi, H. Fan, and Y. Li, \PRA {\bf 91}, 042120 (2015).

\bibitem{gqc} J. I. de Vicente and A. Streltsov, \JPA {\bf 50}, 045301 (2017).
\bibitem{maxcoh} Y. Yao, G. H. Dong, L. Ge, M. Li, and C. P. Sun, \PRA {\bf 94}, 062339 (2016).
\bibitem{coh-pur} A. Streltsov, H. Kampermann, S. W\"{o}lk, M. Gessner, and D. Bru{\ss}, arXiv:1612.07570.
\bibitem{new2} A. E. Rastegin, Front. Phys. {\bf 13}, 130304 (2018).
\bibitem{Bu} K. Bu and C. Xiong, arXiv:1604.06524.
\bibitem{creat3} X. Hu, \PRA {\bf 94}, 012326 (2016).
\bibitem{MUB1} W. K. Wooters, Found. Phys. {\bf 16}, 391 (1986).
\bibitem{MUB2} W. K. Wooters and B. D. Fields, \AoP {\bf 191}, 363 (1989).
\bibitem{cdbas} A. Lakshminarayan, Z. Pucha{\l}a, and K. \.{Z}yczkowski, \PRA {\bf 90}, 032303 (2014).
\bibitem{CHM1} W. Tadej and K. \.{Z}yczkowski, Open Syst. Inf. Dyn. {\bf 13}, 133 (2006).

\bibitem{CHM2} T. Durt, B. G. Englert, I. Bengtsson, and K. \.{Z}yczkowski, Int. J. Quantum Inf. {\bf 8}, 535 (2010).
\bibitem{CHM3} S. Brierley, S. Weigert, and I. Bengtsson, \QIC {\bf 10}, 803 (2010).
\bibitem{peng} Y. Peng, Y. Jiang, and H. Fan, \PRA {\bf 93}, 032326 (2016).
\bibitem{maxrel} C. S. Yu, S. R. Yang, and B. Q. Guo, \QIP {\bf 15}, 3773 (2016).
\bibitem{Dubai} S. Du and Z. Bai, \AoP {\bf 359}, 136 (2015).
\bibitem{Gour} E. Chitambar and G. Gour, \PRA {\bf 94}, 052336 (2016).
\bibitem{Zhuhj} H. Zhu, M. Hayashi, and L. Chen, arXiv:1706.00390.
\bibitem{qubit} H. L. Shi, X. H. Wang, S. Y. Liu, W. L. Yang, Z. Y. Yang, and H. Fan, arXiv:1705.00785.
\bibitem{rqc} M. L. Hu and H. Fan, \PRA {\bf 95}, 052106 (2017).
\bibitem{sud1} M. S. Byrd and N. Khaneja, \PRA {\bf 68}, 062322 (2003).

\bibitem{sud2} G. Kimura, \PLA {\bf 314},339 (2003).
\bibitem{lipur} K. \.{Z}yczkowski, P. Horodecki, A. Sanpera, and M. Lewenstein, \PRA {\bf 58}, 883 (1998).
\bibitem{comple2} U. Singh, M. N. Bera, H. S. Dhar, and A. K. Pati, \PRA {\bf 91}, 052115 (2015).
\bibitem{matrix} R. Bhatia, \textit{Matrix Analysis} (Springer-Verlag, Berlin, 1997).
\bibitem{Haar} F. Mezzadri, Not. Am. Math. Soc. {\bf 54}, 592 (2007).
\bibitem{Haar2} M. L. Mehta, \textit{Random Matrices} (Elsevier, Amsterdam, 2004).


\end{thebibliography}

\end{document}